\documentclass[12pt]{article}
\usepackage{epsfig}

\newcommand\pubnumber{SLAC--PUB--10852}
\newcommand\pubdate{\today}
\newcommand\hepnumber{hep-ph/0411123}

\def\SLAC{Stanford Linear Accelerator Center\\
    Stanford University, Stanford, California 94309, USA}
\def\doeack{\footnote{Work supported in part by the Department of Energy,
                     contract DE--AC02--76SF00515.}}
\def\Graz{Institut f\"ur Physik, Universit\"at Graz \\
Universit\"atsplatz 5, A-8010 Graz,
 Austria}

\def\Title#1{\begin{center} {\Large #1 } \end{center}}
\def\Author#1{\begin{center}{ #1} \end{center}}
\def\Address#1{\begin{center}{ \it #1} \end{center}}
\def\andauth{\begin{center}{and} \end{center}}

\newcommand\pubblock{\rightline{\begin{tabular}{l} \pubnumber\\
         \pubdate \\ \hepnumber \end{tabular}}}
\newenvironment{Abstract}{\begin{quotation} \begin{center}
                       ABSTRACT
     \end{center}\bigskip  }{\end{quotation}}

\def \be  {\begin{equation}}
\def \ee  {\end{equation}}
\def \ba  {\begin{eqnarray}}
\def \ea  {\end{eqnarray}}
\def \baa {\begin{eqnarray*}}
\def \eaa {\end{eqnarray*}}
\def \bb  {\begin{thebibliography}}
\def \eb  {\end{thebibliography}}

\def \as {\relax\ifmmode\alpha_s\else{$\alpha_s${ }}\fi}

\def\sB{\mbox{\scriptsize B}}

\def\sV{\mbox{\scriptsize V}}

\newcommand{\hs}{\hat{s}}
\newcommand{\htt}{\hat{t}}
\newcommand{\hu}{\hat{u}}
\def \Bt {\mbox{B}_{10}}
\def \sBt {{\mbox{\scriptsize B}_{10}}}

\begin{document}
\begin{titlepage}
\pubblock

\vfill \Title{\bf Exclusive Decuplet-Baryon Pair Production \vspace*{1.5mm} \\
 in Two-Photon
Collisions}
\vfill
\Author{ Carola F.~Berger\doeack}
\Address{\SLAC}
\andauth
\Author{ Wolfgang Schweiger}
\Address{\Graz}
\vfill
\begin{Abstract}
This work extends our previous studies of two-photon annihilation
into baryon-antibaryon pairs from spin-1/2 octet to spin-3/2
decuplet baryons.  Our approach is based on perturbative QCD and
treats baryons as quark-diquark systems. Using the same model
parameters as in our previous work, supplemented by QCD sum-rule
results for decuplet baryon wave functions, we are able to give
absolute predictions for decuplet baryon cross sections without
introducing new parameters. We find that the $\Delta^{++}$ cross
section is  of the same order of magnitude as the proton cross
section, well within experimental bounds.
\end{Abstract}
\vfill
\vfill
\end{titlepage}
\def\thefootnote{\fnsymbol{footnote}}
\setcounter{footnote}{0}

\section{Introduction}

The study of exclusive processes in quantum chromodynamics (QCD),
where intact hadrons are explicitly measured in the final state,
provides important insights into the mechanisms of confinement and
into the dynamics of hadronic bound states
\cite{Brodsky:2003dq,Brodsky:2004wx}. Among the multitude of
exclusive processes, two-photon annihilation into
baryon-antibaryon pairs is particularly interesting, because it is
one of the simplest calculable large-angle hadronic scattering
reactions involving two hadrons.  Therefore, $\gamma \gamma
\rightarrow \mbox{B} \bar{\mbox{B}}$ has recently received
considerable experimental \cite{exp} and theoretical
\cite{theor,Berger:2002vc,Karliner:2001ut} attention.

In a recent
paper \cite{Berger:2002vc} we have studied baryon pair production
in two-photon collisions for baryons belonging
to the lowest-lying flavor octet.  In the present note we extend
our work to reactions involving spin-$3/2$ decuplet baryons.
Previously, two-photon annihilation into decuplet baryons has been
studied in Refs.
\cite{Farrar:gv,Anselmino:1987gu,Farrar:1988vz,Karliner:2001ut,
Karliner:2002nk} within different frameworks with
differing conclusions. Thus an experimental analysis could shed
light on the relative importance of the underlying mechanisms
considered here and in the aforementioned references.

Our model  is a modification of the perturbative
hard-scattering picture (HSP) for exclusive processes
\cite{Lepage:1980fj,Efremov:1979qk}.  While the HSP is exactly
valid only at asymptotically large momentum transfer, the
interplay of perturbatively calculable with nonperturbative
effects renders theoretical analyses quite intricate at energies
where data are currently available.  In order to parameterize such
possible non-perturbative effects within a perturbative framework,
an effective formalism was developed in Ref.~\cite{Anselmino:1987vk},
where baryons are treated
as quark-diquark systems.  In the sequel this model has been
successfully applied to a variety of exclusive reactions
\cite{Berger:2002vc,Anselmino:1987gu,Jakob:1993th,prediqu1,prediqu2,
Berger:1999gx}.

In the following, we start with a brief review of the
quark-diquark model.  Then we go on to describe the new
ingredients necessary for the study of processes involving
decuplet baryons. In Sec.~\ref{sec:results} we present and discuss
model predictions with emphasis on the $\Delta$ cross sections,
for which experimental upper bounds are available~\cite{Argus}.
Following concluding remarks, supplementary analytical expressions
for the scattering amplitudes are tabulated in the Appendix.

\section{Exclusive Reactions in the \\ Quark-Diquark Picture}

Here we briefly summarize the modified hard-scattering formalism with
diquarks, and elaborate on the aspects specific to the treatment of
decuplet baryons. For a full account of all details we refer to our
recent work \cite{Berger:2002vc,Berger:1999gx}.

\subsection{Review of the Model}\label{sec:hsp}

As in the conventional hard-scattering picture
\cite{Lepage:1980fj,Efremov:1979qk}, an exclusive reaction amplitude
$\mathcal{M}$ is convolutively factorized into a process-dependent,
perturbative hard-scattering amplitude $\hat{T}$ and
process-independent, non-perturbative distribution amplitudes $\Psi$.
The latter are probability amplitudes for finding the pertinent
valence Fock states, here quarks and diquarks, in the scattering
hadrons.  The amplitude for two-photon annihilation into a
baryon-antibaryon pair is given by
\begin{eqnarray}
\overline{{\mathcal{M}}}_{\{\lambda\} }\! \left(\hs, \htt\right)
\! & = & \!\!\! \int\limits_0^1 \!\! d x_1 \!\! \int\limits_0^1
\!\! d y_1 \Psi_{\sB}^\dagger \left(x_1\right)
\Psi_{\overline{\sB}}^\dagger \left(y_1\right)
\hat{T}_{\{\lambda\}\!\! } \left(x_1, y_1; \hs, \htt\right),
\nonumber \\  \label{HSP}
\end{eqnarray}
where Lorentz and color indices are suppressed for convenience.
Furthermore,
the dependence on renormalization and factorization scales is
neglected since we are only interested in a rather restricted  range
of momentum transfer.  The subscript $\left\{ \lambda \right\}$
denotes all possible configurations of photon and baryon helicities.
In the following we use the label $\mathrm{B}$ to denote spin-1/2
octet baryons and $\Bt$ to label spin-3/2 decuplet baryons.

For the process $\gamma \gamma \rightarrow \Bt
\overline{\mbox{B}}_{10}$, there are 19 independent helicity amplitudes,
$\overline{\mathcal{M}}_{\lambda_{\sBt},\,\lambda_{\overline{\sB}_{10}};
\,\lambda_1,\,\lambda_2}$, where the
$\lambda_{\sBt},\,\lambda_{\overline{\sB}_{10}}$ are the helicities of
the outgoing baryon and antibaryon, respectively, and
$\lambda_{1},\,\lambda_{2}$ label the helicities of the two photons.
Only 13 out of these 19 helicity amplitudes involve a zero or single
flip of the hadronic helicity. Double flip amplitudes vanish in
our approach. We use the following convention for
the nonvanishing amplitudes:
\begin{samepage}
\begin{equation}
\begin{array}{lcl}
\overline{\phi}_1
 =  \overline{\mathcal{M}}_{-\frac{1}{2},\,\frac{1}{2};\,1,\,-1}, &
 \quad &
\overline{\phi}_7 =
\overline{\mathcal{M}}_{-\frac{1}{2},\,\frac{3}{2};\,1,\,-1}, \\
\overline{\phi}_2  =
\overline{\mathcal{M}}_{-\frac{1}{2},\,-\frac{1}{2};\,1,\,1}, &
 \quad &
\overline{\phi}_8 =
\overline{\mathcal{M}}_{\frac{1}{2},\,-\frac{3}{2};\,1,\,1}, \\
\overline{\phi}_3  =
\overline{\mathcal{M}}_{\frac{1}{2},\,-\frac{1}{2};\,1,\,1}, &
 \quad &
\overline{\phi}_9 =
\overline{\mathcal{M}}_{\frac{1}{2},\,-\frac{3}{2};\,1,\,-1}, \\
\overline{\phi}_4  =
\overline{\mathcal{M}}_{\frac{1}{2},\,\frac{1}{2};\,1,\,-1}, &
 \quad &
\overline{\phi}_{10} =
\overline{\mathcal{M}}_{-\frac{1}{2},\,\frac{3}{2};\,1,\,1}, \\
\overline{\phi}_5  =
\overline{\mathcal{M}}_{\frac{1}{2},\,-\frac{1}{2};\,1,\,-1}, &
 \quad &
\overline{\phi}_{11} =
\overline{\mathcal{M}}_{-\frac{3}{2},\,\frac{3}{2};\,1,-\,1}, \\
\overline{\phi}_6  =
\overline{\mathcal{M}}_{\frac{1}{2},\,\frac{1}{2};\,1,\,1}, &
 \quad &
 \overline{\phi}_{12} =
\overline{\mathcal{M}}_{\frac{3}{2},\,-\frac{3}{2};\,1,\,1}, \\
 & \quad &
\overline{\phi}_{13} =
\overline{\mathcal{M}}_{\frac{3}{2},\,-\frac{3}{2};\,1,\,-1}.
\label{annamps}
\end{array}
\end{equation}
\end{samepage}
Other helicity configurations are related to these via parity and/or time
reversal invariance.  Our normalization of the amplitudes is such that
the differential cross section for two-photon annihilation into
decuplet baryons is given by
\be \frac{d \sigma}{d t} = \frac{1}{64
\pi s^2} \sum_{\left\{\lambda\right\}} \left|
\overline{\mathcal{M}}_{\left\{\lambda\right\}} \right|^2,
\ee
where the sum is over all possible helicity configurations
$\{\lambda\}$.

In (\ref{HSP}), $\hat{T}$ consists of all possible tree diagrams
that contribute to the elementary scattering process $\gamma
\gamma \rightarrow q D \bar{q} \bar{D}$. The momenta carried by
quarks $q$ and diquarks $D$ are assumed to be collinear to those
of their parent hadrons, $\mbox{B}$. The quark and antiquark carry
momentum fractions $x_1$ and $y_1$ in the baryon and antibaryon,
respectively, while the diquark and antidiquark carry momentum
fractions $x_2 = 1-x_1$ and $y_2 = 1-y_1$, respectively. Since we
assume that every baryonic constituent has a four-momentum $x \,
p_{\sB}$ proportional to the four-momentum of its parent hadron
$p_{\sB}$ \cite{Anselmino:vs}, it acquires an effective mass $x
m_{\sB}$, where $m_{\sB}$ denotes the baryon mass. These effective
masses are taken into account for all internal and external legs
of the Feynman diagrams contributing to the hard-scattering
amplitude $\hat{T}$. The hard-scattering amplitude is then
expanded in powers of the small parameter $(m_{\sB}/\sqrt{s} )$ up
to next-to-leading order, at fixed center-of-mass scattering angle
$\hat{\theta}$. The result is reexpressed in terms of massless
Mandelstam variables, $\hs$, $\htt$, and $\hu$ which are obtained
from the usual massive Mandelstam variables, $s,\,t,\,u$, again by
expansion in $(m_{\sB}/\sqrt{\hs} )$. In the hard scattering
diagrams, the composite nature of the diquarks is taken into
account by diquark form factors. These are parameterized such that
asymptotically the scaling behavior of the pure quark HSP emerges.

The complete parameterization of the  model, including form
factors and octet-baryon wave functions can be found in
\cite{Berger:2002vc}. These parameters were fixed in
\cite{Jakob:1993th} by fitting elastic electron-nucleon scattering
data. With the same set of parameters a variety of other processes
has been computed, and the results have successfully met
experimental comparison
\cite{Berger:2002vc,Jakob:1993th,prediqu2,Berger:1999gx}.

\subsection{Decuplet Baryons}\label{decuplet}

The diquark model comprises spin-0 (scalar) and spin-1
(vector) diquarks.  While both scalar (S) and vector (V) diquarks
 contribute to
processes involving spin-$1/2$ octet baryons, the valence Fock states
of spin-$3/2$ decuplet baryons consist only of quarks and vector
diquarks.

We recall that the valence Fock state of an
octet baryon $\mathrm{B}$ with mass $m_{\sB}$, momentum $p_{\sB}$,
and helicity $\lambda$ can be described by the following quark-diquark
wave function
\be
\Psi_{\sB}(p_{\sB}, x, \lambda)  =  f_{S}^{\sB} \Phi_{S}^{\sB}
(x) \chi_{S}^{\sB} \,u(p_{\sB},\lambda)
 +  f_{V}^{\sB} \Phi_{V}^{\sB}(x) \chi_{V}^{\sB} \frac{1}
{ \sqrt{3}}\left( \gamma^\mu + \frac{p_{\sB}^\mu }{m_{\sB} }
\right) \gamma_5\,u(p_{\sB},\lambda) \label{wave8}\ee
when transverse momenta of the constituents are neglected. $x$ is
the longitudinal momentum fraction of the quark, whereas the
diquark carries the longitudinal momentum fraction $1-x$.
Analogously, the wave function of a decuplet baryon may be
written as
\be
\Psi_{\sBt}^\mu (p_{\sBt},x,\lambda) = f_V^{\sBt} \Phi_V^{\sBt}(x)
\chi_V^{\sBt} u^\mu\left(p_{\sBt},\lambda\right), \label{wave}
\ee
with the Rarita-Schwinger spinors \cite{Rarita:mf}
\ba u^\mu
\left(p,\lambda = \pm 3/2 \right) & = & \varepsilon^\mu\left(\pm 1
\right) \,u\left(p,\lambda = \pm 1/2 \right) \, ,\nonumber \\
u^\mu\left(p,\lambda = \pm 1/2 \right) & = & \left[\sqrt{\frac{3}{2}}
\varepsilon^\mu(0)  - \frac{2 \lambda}{\sqrt{6}} \left( \gamma^\mu +
\frac{p^\mu}{m_{\sBt}} \right) \gamma_5 \right] u(p, \lambda).
\ea
Recall that all Lorentz indices have been suppressed in
Eq.~(\ref{HSP}), the open index $\mu$ of the vector diquark
polarization vector in (\ref{wave8}) and (\ref{wave}) is
contracted appropriately in the convolution integral (\ref{HSP}).
$\chi_D^{\sB}$, $\chi_D^{\sBt}$ ($D=S,V$) denote pertinent SU(3)
quark-diquark flavor wave functions and $\Phi_D^{\sB}$,
$\Phi_D^{\sBt}$ represent the nonperturbative probability
amplitudes for finding these constituents with momentum fractions
$x$ and $1-x$, respectively, in the (decuplet) baryon. These
probability amplitudes are normalized such that
\be
\int\limits_0^1 dx \Phi_D^{\sB}(x) = 1,
\label{DAnorm}
\ee
and analogously for $\Phi_D^{\sBt}$. The constants $f_D^{\sB}$,
$f_D^{\sBt}$ result from integrating out intrinsic transverse
momenta in the full wave function to produce Eqs. (\ref{wave8}) and
(\ref{wave}), respectively. The numerical values of $f_D^{\sB}$
and $f_D^{\sBt}$ are furthermore
determined by the overall probability of
finding the $|qD\rangle$-state in the baryon $\mathrm{B}$ or
decuplet baryon $\Bt$, respectively.

For unbroken SU(6) spin-flavor symmetry octet- and decuplet baryon
wave functions are related, specifically,
$\Phi_{S}^{\sB}=\Phi_{V}^{\sB}=\Phi_{V}^{\sB_{10}}$ and
$f_{S}^{\sB}=f_{V}^{\sB}=f_{V}^{\sB_{10}}/\sqrt{2}$.  In the
actual parameterization of the diquark model~\cite{Jakob:1993th}
the asymptotic SU(6) symmetry is systematically broken down to
SU(3) flavor symmetry. Thus the above SU(6) relations are by no
means satisfied, and $\Phi_{S}^{\sB}$ and $\Phi_{V}^{\sB}$ as well
as $f_{S}^{\sB}$ and $f_{V}^{\sB}$ have quite different values.
Since SU(6) symmetry is thus already broken within the baryon
octet we cannot use SU(6) symmetry for deriving quark-diquark wave
functions of decuplet baryons. Instead, we will apply another
strategy to fix $\Phi_{V}^{\sB_{10}}$ and $f_{V}^{\sB_{10}}$.

 The lowest moments of
three-quark wave functions of octet and decuplet baryons
are restricted by QCD sum
rules~\cite{Farrar:1988vz,Chernyak:1989}.  Model wave functions
that satisfy the QCD sum-rule constraints (for a typical
factorization scale of about $1$~GeV) are very asymmetric in the
longitudinal momentum fractions $x_i,\,i=1,2,3$ of the quarks for
octet baryons and nearly symmetric ($\sim x_1 x_2 x_3$) for the
$\Delta$s and the $\Omega$~\cite{Farrar:1988vz}.  By regrouping
terms in the three-quark wave function such that, for example,
quarks 2 and 3 are in a specific spin-flavor state and by
integrating over one of the momentum fractions of the two quarks
that build up this \lq\lq diquark\rq\rq\ we can convert the
three-quark wave function into a quark-diquark wave function that
nearly has the form (\ref{wave8}) or (\ref{wave}) for octet or
decuplet baryons, respectively. For more information on this
conversion we refer to \cite{Anselmino:1987gu}. The probability
amplitudes $\Phi_{V}^{\sB}$ and $\Phi_{V}^{\sBt}$ for general
three-quark wave functions are different in the cases of
helicity-0 and helicity-1 V diquarks. For the octet and decuplet
model wave functions that we employ this difference turns out to
negligible. We then arrive at
 Eq.~(\ref{wave8}) or (\ref{wave}), respectively.

We apply the above procedure  to the three-quark
wave function of the $\Delta$ that has been proposed in
Ref.~\cite{Farrar:1988vz} based on QCD sum-rule constraints.
We obtain the following quark-diquark wave function
for a $\Delta$ with helicity $\pm1/2$
\be \Phi_{V}^{\Delta, |\lambda|=1/2}(x) = N x (1-x)^3 (1 - 2.95 x
+ 3.86 x^2) \exp \left\{-b^2 \left[ \frac{m_q^2}{x} + \frac{m_V^2
}{1-x} \right] \right\} \, . \label{huangfull} \ee
Analogous to the standard parameterization of the diquark model
for octet baryons~\cite{Jakob:1993th} we have introduced an
additional exponential factor that damps the end-point regions
$x\rightarrow 0,1$.  Such an exponential factor results if the
transverse momentum dependence  of the full wave function, which is
integrated over, is assumed to be of Gaussian form.  The
parameters $b^2=0.248$~GeV$^2$, $m_q=0.33$~GeV, and $m_V=0.58$~GeV
are taken to be the same as for octet baryons. The normalization
factor $N$ is determined by Eq. (\ref{DAnorm}). The expression for
$\Phi_{V}^{\Delta, |\lambda|=3/2}(x)$ differs, in general, from
$\Phi_{V}^{\Delta, |\lambda|=1/2}(x)$.  However, we refrain from
quoting it here, because our explicit calculations show that the
production of helicity-3/2 $\Delta$s is suppressed within the
diquark model.

The only remaining open parameter is now
the normalization $f_{V}^{\Delta,|\lambda|=1/2}$
of the helicity-$1/2$ $\Delta$ wave function.  Since the
normalization $f_{V}^{\sB}$ of the octet-baryon wave function was
taken as a free parameter in the diquark model we normalize the
$\Delta$ wave function relative to the proton wave function.  This
means that we convert the three-quark wave functions for proton
and $\Delta$ into quark-diquark wave functions of the form
(\ref{wave8}) and (\ref{wave}), respectively, and consider the
resulting ratio $f_{V}^{\Delta,|\lambda|=1/2}/f_{V}^{\mathrm{p}}$.
For the QCD sum-rule based wave functions of Refs.~\cite{Farrar:1988vz}
and \cite{Chernyak:1989} this ratio becomes
$f_{V}^{\Delta,|\lambda|=1/2}/f_{V}^{\mathrm{p}}=0.898$.  With
$f_{V}^{\mathrm{p}}=127.7$~MeV, the value obtained in a fit of
elastic electron-nucleon scattering data~\cite{Jakob:1993th}, we
thus find
\be f_{V}^{\Delta,\,|\lambda|= 1/2} =  125.1  \mbox{ MeV} \, .
\label{fVD} \ee

This completes the parameterization of our model for decuplet
baryons. For sake of completeness we quote the flavor wave
functions entering (\ref{wave}) for the differently charged
$\Delta$s:
%
\ba \chi_V^{\Delta^{++}} &=& u V_{\{uu\}} \, ,\nonumber\\
\chi_V^{\Delta^{+\phantom{+}}} &=& \left[
\sqrt{2}u V_{\{ud\}}+d V_{\{uu\}}\right]/\sqrt{3}\, , \nonumber \\
\chi_V^{\Delta^{0\phantom{+}}} &=& \left[
\sqrt{2}d V_{\{ud\}}+u V_{\{dd\}}\right]/\sqrt{3}\, ,\nonumber \\
\chi_V^{\Delta^{-\phantom{+}}} &=& d V_{\{dd\}}\, . \label{flavorwf}\ea

\section{Results} \label{sec:results}

We list analytical results for the hard-scattering amplitudes
$\hat{T}_{\{\lambda\} }$ contributing to $\gamma \gamma
\rightarrow \Bt \bar{\mbox{B}}_{10}$ in the Appendix.  These
results have been checked via crossing relations
\cite{Bourrely:mr} against the separately computed amplitudes for
the crossed process, Compton scattering $\gamma \Bt \rightarrow
\gamma \Bt$. Comparing the spinor structure of the decuplet baryon
wave function (\ref{wave}) with the one for octet baryons
(\ref{wave8}), we find that the leading, non-flip, hard amplitudes
for decuplet baryons with helicity 1/2 are related by a factor of
2 to those for octet baryons.  From the analytical expressions we
also  observe, that the hard-scattering amplitudes for octet
baryons with helicity $\pm 3/2$ are suppressed by
${\mathcal{O}}(m_{\sBt}^2/\hs)$ or higher, even if these
amplitudes conserve the hadronic helicity or flip it by one unit.
The only 4-point contribution that is not suppressed enters the
helicity amplitude $\bar{\phi}_{2}$.  In the numerical
calculations this contribution, however, turns out to be nearly
negligible. 5-point functions with both photons attaching to the
diquark do not contribute at all, since these are also suppressed
by ${\mathcal{O}}(m_{\sBt}^2/\hs)$ or even higher.

\begin{figure}[h!]
\begin{center}
\epsfig{file=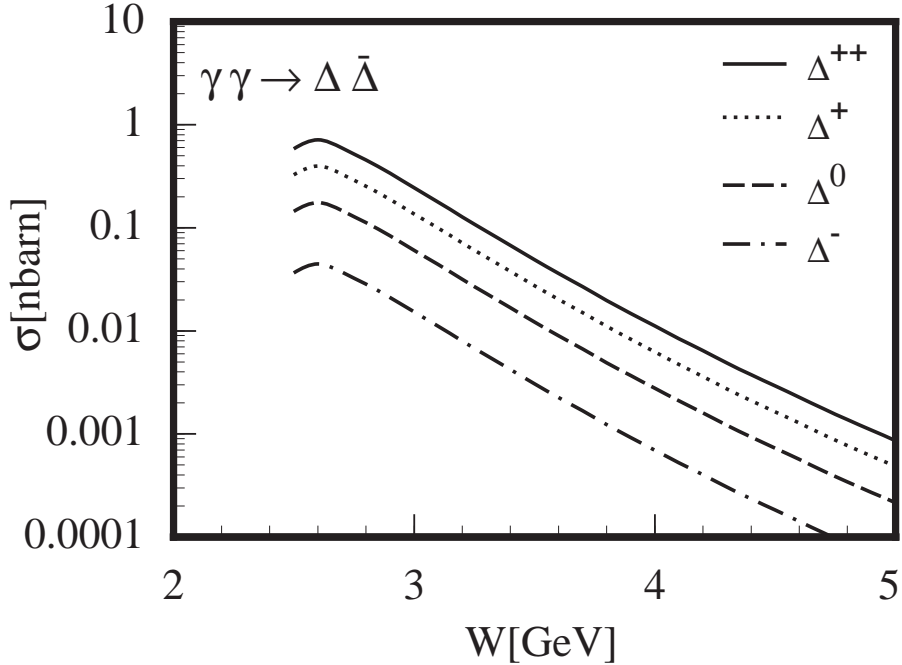,width=12cm,angle=0,clip=0} \caption{
Integrated cross sections for $\gamma \gamma \rightarrow
\Delta^{++} \bar{\Delta}^{--}$ (solid line), $\Delta^{+}
\bar{\Delta}^{-}$ (dotted), $\Delta^{0} \bar{\Delta}^{0}$
(dashed), $\Delta^{-} \bar{\Delta}^{+}$ (dash-dotted line)
($|\cos(\theta_{CM})|<0.6$) versus center-of-mass energy $W =
\sqrt{s}$ predicted with the standard parameterization of the
diquark model~\cite{Berger:2002vc} and the $\Delta$ DA defined in
the text (see Eqs.~(\ref{wave}), (\ref{huangfull}), and
(\ref{fVD})). \label{fig:deltas}}
\end{center}
\end{figure}

As a consequence of this observation, cross section ratios of different
decuplet baryon channels can easily be estimated, provided that the
corresponding probability amplitudes $\Phi_V^{\sBt}$ are not too
different.  The cross-section ratios are then essentially determined
by the corresponding charge-flavor factors $C_{\mathrm{cf}}^{(3)}$
(see Eq.(\ref{hsa})) and the wave function normalizations
$f_V^{\sBt}$.  For the $\Delta$-quartet $\Phi_V^{\Delta}$ and
$f_V^{\Delta}$ are the same for all members due to isospin symmetry.
From the flavor wave functions (\ref{flavorwf}) the
charge-flavor factors $C_{\mathrm{cf}}^{(3)}$ are seen to be $4/9$,
$3/9$, $2/9$, and $1/9$ for the $\Delta^{++}$, $\Delta^{+}$,
$\Delta^{0}$, and $\Delta^{-}$, respectively. The cross section
ratios become (approximately)
\be \sigma(\Delta^{++}): \sigma(\Delta^{+}): \sigma(\Delta^{0}):
\sigma(\Delta^{-}) = 16 : 9 : 4 : 1 \, . \label{ratios}\ee
This is the first interesting prediction of the diquark model.  In
Fig.~\ref{fig:deltas} we show the integrated cross sections
($|\cos(\theta_{CM})|<0.6$, where $\theta_{CM}$ is the center-of-mass
scattering angle) for the $\Delta$ channels.  The plot exhibits
numerical predictions obtained with the standard parameterization of
the diquark model~\cite{Berger:2002vc} and the $\Delta$ wave function
derived in Sec.~\ref{decuplet}.  It confirms Eq.~(\ref{ratios})
within 1 percent.

This prediction is to be contrasted with the ratios $16:1:0:1$
that result if the photons couple to the total charge of the
$\Delta$s. Also within the pure quark HSP the ratios for the
$\Delta^{+}$ and the $\Delta^{0}$ channels differ from ours.
Within the pure quark HSP the cross section ratios for the
different $\Delta$ channels are predicted to be
$\sigma(\Delta^{++}): \sigma(\Delta^{+}): \sigma(\Delta^{0}):
\sigma(\Delta^{-}) \approx 16 : 2 : 1/3 : 1
$~\cite{Farrar:1988vz}. Note that all the above predictions agree
with our result for the cross section ratio $\sigma(\Delta^{++}) :
\sigma(\Delta^{-}) \approx 16:1 $. This result is also found
in a
more general QCD analysis~\cite{Karliner:2002nk}. However, yet another
possible production mechanism via multi-pion intermediate states
predicts $\sigma(\Delta^{++})= \sigma(\Delta^{-})$ and
$\sigma(\Delta^{+})= \sigma(\Delta^{0})$~\cite{Karliner:2001ut}.
An experimental determination of such cross section ratios could
therefore provide important clues on the underlying production
mechanisms, especially because in ratios of cross sections for
different $\Delta$ channels the sensitivity to the specific form
of the $\Delta$ wave function should be greatly reduced.

If we assume SU(3)-flavor symmetry, that is, if we take the same
$\Phi_V^{\sBt}$ and $f_V^{\sBt}$ for all decuplet baryons, we are
also able to give estimates for the pair production of strange
decuplet baryons.  Aside from appropriate phase space factors,
SU(3) symmetry implies
\ba
\sigma(\Delta^{+})
& = & \sigma(\Sigma^{\ast +}) \, , \nonumber \\
\sigma(\Delta^{0}) & = & \sigma(\Sigma^{\ast 0})  = \sigma(\Xi^{\ast 0})\, , \nonumber \\
\sigma(\Delta^{-}) & = & \sigma(\Sigma^{\ast -}) = \sigma(\Xi^{\ast -}) =
\sigma(\Omega^{\ast -}) \, .
\ea
However, since it is experimentally very
difficult to measure pair-production cross
sections for decuplet baryons, we refrain from giving
quantitative results for the strange decuplet baryons.  We
rather concentrate in the following on the
$\Delta^{++}$ channel which might have the best chance to be measured
due to its comparably large cross section.

\begin{figure}[h!]
\begin{center}
\epsfig{file=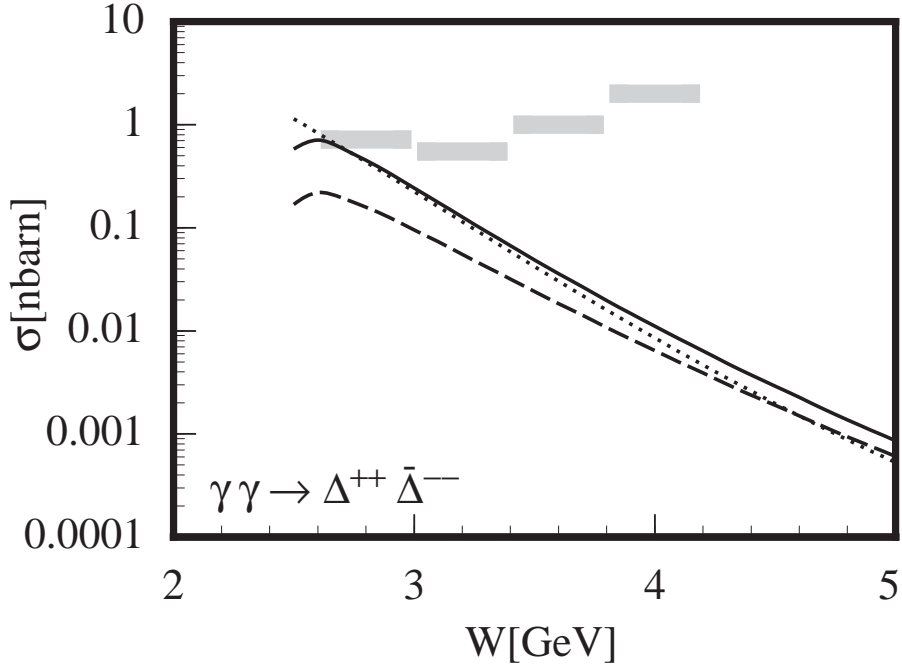,width=12cm,angle=0,clip=0} \caption{
Integrated cross section for $\gamma \gamma \rightarrow
\Delta^{++} \bar{\Delta}^{--}$ ($|\cos(\theta_{CM})|<0.6$) versus
$W = \sqrt{s}$ for the same $\Delta$ DA as in
Fig~\ref{fig:deltapp}. The solid line corresponds to the full
diquark-model calculation. The contribution to the cross section
that comes from the hadronic-helicity conserving amplitudes
$\bar{\phi}_{1}$ and $\bar{\phi}_{5}$ is represented by the dashed
line.  For comparison we also show the integrated cross section
for $\gamma \gamma \rightarrow \mathrm{p} \bar{\mathrm{p}}$
(dotted line) calculated within the same
model~\cite{Berger:2002vc}.  The shaded boxes indicate
experimental upper bounds as obtained by the ARGUS
collaboration~\cite{Argus}.\label{fig:deltapp} }
\end{center}
\end{figure}

In Fig.~\ref{fig:deltapp} we show for comparison with the $\gamma
\gamma \rightarrow \Delta^{++} \bar{\Delta}^{--}$ cross section the
$\gamma \gamma \rightarrow \mathrm{p} \bar{\mathrm{p}}$ cross section
that we have obtained with the same
parameterization~\cite{Berger:2002vc}.  Surprisingly, we find that the
$\Delta^{++}$ cross section is of the same order of magnitude as the
proton cross section.  This prediction seems to be very stable against
(reasonable) changes of the $\Delta$ wave function.  With a $\Delta$
wave function that satisfies the SU(6) relations
$\Phi_{V}^{\Delta}=\Phi_{V}^{\mathrm{p}}$ and $f_{V}^{\Delta}=\sqrt{2}
f_{V}^{\mathrm{p}}$ we obtain, for example, a result which is only
about 20\% to 30\% smaller
\footnote{In a previous attempt to estimate
$\sigma(\Delta^{++})/\sigma(\mathrm{p})$ within a diquark model a
ratio of $\approx 0.1$ was found~\cite{Anselmino:1987gu}.  This,
however, was obtained with an incomplete version of the diquark model,
where V diquarks were not taken fully into account and mass
effects have been neglected.}.
Under the naive assumption that the
photons couple directly to the charges of the baryons one would expect
the $\Delta^{++}$ cross section to be about 16 times larger than the
proton cross section.

From the viewpoint of the pure quark hard-scattering picture, the
ratio of $\gamma \gamma \rightarrow \Delta^{++} \bar{\Delta}^{--}$ to
the $\gamma \gamma \rightarrow \mathrm{p} \bar{\mathrm{p}}$ cross
section depends strongly on the choice of the proton wave
function~\cite{Farrar:1988vz}.  Not surprisingly, a result for the
ratio comparable to ours is obtained with the QCD sum-rule wave
functions of Refs.~\cite{Farrar:1988vz} and \cite{Chernyak:1989} for
$\Delta$ and proton, respectively, which we have used in
Sec.~\ref{decuplet} to derive and normalize our quark-diquark wave
function of the $\Delta$.  However, if the asymptotic wave function
$\sim x_1 x_2 x_3$ is taken for both the proton and the $\Delta$,
the cross section ratio $\sigma(\Delta^{++})/\sigma(\mathrm{p})$ can
be as large as 50 within the pure quark HSP~\cite{Farrar:gv}.  On the
other hand, soliton models involving multi-pion channels predict a
much smaller ratio \cite{Brodsky:2004wx,Karliner:2001ut}, comparable
to our findings.  An experimental determination of the ratio
$\sigma(\Delta^{++})/\sigma(\mathrm{p})$ could therefore help to
explore the importance of the various mechanisms that result in these
quite different predictions.

Unfortunately, it is very difficult experimentally to isolate the
signal of the broad $\Delta^{++}$ resonance from the background
and to disentangle the $\Delta^{++}$ and the $\Delta^{0}$
contributions in the $\gamma \gamma \rightarrow \mathrm{p}
\bar{\mathrm{p}} \pi^+ \pi^-$ cross sections which are actually
measured. Therefore only upper limits for the $\gamma \gamma
\rightarrow \Delta^{++} \bar{\Delta}^{--}$ cross section have been
extracted up to now by the ARGUS collaboration~\cite{Argus}. As
can be seen in Fig. \ref{fig:deltapp}, our results lie well below
these upper limits. More recent attempts to constrain the $\gamma
\gamma \rightarrow \Delta^{++} \bar{\Delta}^{--}$ cross section
using the data taken by the L3 group are afflicted with the same
problems, but a preliminary assessment indicates compatibility
with the ARGUS
results and our predictions~\cite{Echenard}. A
better chance to determine the cross section for
 $\Delta^{++} \bar{\Delta}^{--}$
pair production would perhaps exist for the BABAR or BELLE
experiments which enjoy a much higher luminosity.

Finally, let us comment on the treatment of mass effects within
our approach. Fig.~\ref{fig:deltapp} displays the effect of taking
into account the finite $\Delta$ mass.  As explained in
Sec.~\ref{sec:hsp}, the $\Delta$ mass is taken into account
in the hard-scattering amplitudes via an expansion in the
small parameter $(m_{\sB}/\sqrt{\hs} )$ where only the leading and
next-to leading order terms are kept.  As expected, mass
correction terms do not contribute to the hadronic helicity-conserving
amplitudes $\overline{\phi}_1$ and $\overline{\phi}_5$.  Only the
amplitudes that involve a single flip of the hadronic helicity, which
vanish if masses are neglected, become nonzero due to the mass
correction terms. The comparison of the solid and the dashed lines in
Fig.~\ref{fig:deltapp} shows that these mass effects can be sizable in
the few-GeV region.  At $W=2.5$~GeV the leading-order contributions
provide only about 30\% of the full cross section.  This ratio
increases, of course, with increasing energy and becomes
roughly 70\% at $W=5$~GeV.

\section{Concluding Remarks}

In this work we have computed $\gamma \gamma \rightarrow \mbox{B}_{10}
\bar{\mbox{B}}_{10}$ cross sections at intermediate momentum
transfer for the case of spin-3/2 decuplet baryons
$\mbox{B}_{10}$. We have employed a modification of the
hard-scattering picture for exclusive reactions, where baryons
are treated as quark-diquark systems, thereby effectively
parameterizing nonperturbative contributions which are undoubtedly
present at currently experimentally accessible energies. Using the
same model parameters as in previous studies
 of other photon-induced
 reactions, and constraining the quark-diquark wave
function of the $\Delta$ with the help of QCD sum-rule results,
we are able to give absolute predictions
for $\gamma \gamma \rightarrow \Delta \bar{\Delta}$ without
introducing new parameters.

We find that the cross section for $\gamma \gamma \rightarrow
\Delta^{++} \bar{\Delta}^{--}$ is of the same order of magnitude
as the cross section for proton pair production, $\gamma \gamma
\rightarrow p \bar{p}$. Furthermore, we observe that
the pair production of decuplet baryons is almost completely
determined within our model
by those graphs where both photons couple to the
quark line. This enables us to estimate production
ratios for different decuplet-baryon channels independent of the
choice of the wave function, provided that the wave functions are
similar for all baryons within the decuplet.
This is certainly the case for
the $\Delta$-quartet for which we predict the ratios
$\sigma(\Delta^{++}):
\sigma(\Delta^{+}): \sigma(\Delta^{0}): \sigma(\Delta^{-}) = 16 :
9 : 4 : 1$.

There are various other estimates of these cross section ratios
in the literature, based on
different viewpoints and production mechanisms, which differ in
their predictions from ours. It would therefore be necessary to
compare to experimental analyses, in order to determine the
relative importance of the considered production mechanisms, and
to learn more about the degree of symmetry among constituents in
decuplet-baryon distribution amplitudes. Such an experimental
analysis should be quite feasible at a
high-luminosity $e^+e^-$ collider. We therefore hope that our
experimental colleagues will study this
interesting problem in the near future.

\begin{appendix}
\section{Elementary Helicity Amplitudes \\ for $\gamma \gamma
\rightarrow q V \bar{q} \bar{V}$}

There are 30 Feynman graphs that contribute to the hard-scattering
amplitudes $\hat{T}$ for $\gamma \gamma \rightarrow q V \bar{q}
\bar{V}$. Their general structure is
\be \hat{T}_{\{\lambda\} }(\htt,\hu) =   C_{\mathrm{cf}}^{(3)}
\,\overline{T}^{(3, V)}_i (\htt,\hu) F_V^{(3)} +
C_{\mathrm{cf}}^{(4)} \,\overline{T}^{(4, D)}_i(\htt,\hu)
F_V^{(4)} + C_{\mathrm{cf}}^{(5)}\, \overline{T}^{(5, V)}_i
(\htt,\hu) F_V^{(5)}, \label{hsa}\ee
where $C_{\mathrm{cf}}^{(n)}$ are the appropriate charge-flavor
factors. The subscript $i = 1,\dots, 13$ labels the
helicity-combinations according to Eq.~(\ref{annamps}). Each
$n$-point contribution $\overline{T}^{(n,V)}$ is found from a
separately gauge-invariant set of Feynman diagrams, where $(n-2)$
gauge bosons couple to the diquark. The $\overline{T}^{(n,V)}$
are multiplied with the appropriate diquark form factors
$F_V^{(n)}$, parameterizing the composite nature of diquarks.
For further details we refer to
\cite{Berger:2002vc}.

The analytical results for $\overline{T}^{(n,V)}_i$ are presented
in the following. For their calculation we employed the algebraic
computer program \textit{Mathematica}~\cite{math} with the package
\texttt{FeynCalc}~\cite{Mertig:1990an}. We do not list $n$-point
contributions that are suppressed by at least
${\mathcal{O}}(m_{\sBt}^2/\hs)$, since they are neglected in our
numerical calculations. Those include, for example, all 5-point
functions $\overline{T}^{(5, V)}_i$ and all amplitudes with
(anti)baryon helicity $\pm \frac{3}{2}$. Note that the
parameterization of the form factors $F_V^{(n)}$ for 4- and
5-point functions provides additional inverse powers of $\hs$ as
compared to $F_V^{(3)}$. We abbreviate $C = (4 \pi)^2 C_F \alpha
\,\alpha_s$, where $C_F = \frac{4}{3}$ is the color factor, and
$\alpha$ denotes the fine structure constant $\alpha \approx
1/137$. $\kappa_{\sV}$ is the anomalous magnetic moment of the
vector diquark.
\begin{eqnarray*}
 \overline{T}_1^{(3,V)} (\htt,\hu) & = & -\frac{4}{3} C
\frac{\kappa_{\sV}}{m_{\sBt}^2
\sqrt{\hu \htt }} \left( \frac{\hu }{x_1 y_1 } + \frac{\htt }{x_2
y_2} \right)
 \\
\overline{T}_2^{(3,V)} (\htt,\hu) & = & \frac{2}{3} C
\frac{1}{m_{\sBt} }
\frac{ \sqrt{\hs} \,\hs}{\hu \htt } \frac{x_1 + y_1 }{x_1 y_1 } \\
\overline{T}_4^{(3,V)} (\htt,\hu) & = & C\frac{2 }{3\, m_{\sBt}  }
\frac{1}{\sqrt{\hs}\,\hu \htt} \frac{1 }{x_1 x_2 y_1 y_2 } \Bigg\{   \\
& &
- \kappa_{\sV} \left[ (2 x_1 -3) y_2 \htt^2 + (2 y_1 - 3) x_2 \hu^2
-   4 x_1 y_1 \hu \htt
   \right]+ \\
& & + \,\left[ \left( x_2 + y_2 \right) \left( \htt^2 x_1 y_2 + \hu^2
x_2 y_1 - 2 x_1 y_2 \hu \htt \right)    - y_2 \htt^2 - x_2 \hu^2
\right]
\Bigg\} \\
\overline{T}_5^{(3,V)} (\htt,\hu) & = & \overline{T}_1^{(3,V)}
(\hu,\htt) \\
  \overline{T}_6^{(3,V)} (\htt,\hu) & = & - C \frac{2}{3} \frac{1
}{m_{\sBt}}
\frac{ \sqrt{\hs} \, \hs}{\hu \htt } \left[ - (1 + \kappa_{\sV})
\frac{x_1 y_2^2+ y_1 x_2^2 }{x_1 x_2 y_1 y_2 }
 + \kappa_{\sV} \frac{   x_1 + y_1}{x_2 y_2 }  \right] \\
\overline{T}_2^{(4,V)} (\htt,\hu) & = &   - \frac{2}{3} C \frac{
\kappa_{\sV}
 (1 - \kappa_{\sV}) \sqrt{\hs}}{m_{\sBt}^3} \frac{1}{x_1 x_2^2 y_1
y_2^2 }.
\end{eqnarray*}

The hard-scattering amplitudes for Compton scattering off decuplet
baryons are related to the amplitudes listed above via crossing
\cite{Bourrely:mr}. The corresponding elementary helicity
amplitudes, $\gamma q D \rightarrow \gamma q D$ have been computed
separately as a check.
They can be obtained from the authors upon request.

\end{appendix}

\subsection*{Acknowledgements}
We thank Mauro Anselmino, Mariaelena Boglione, Stan Brodsky, Lance
Dixon, and Marek Karliner for stimulating and helpful discussions.
We are also grateful to Bertrand Eche\-nard for sharing his
insights from an experimental viewpoint.


\begin{thebibliography}{99}

\bibitem{Brodsky:2003dq}
M.~R.~Pennington,
Nucl.\ Phys.\ Proc.\ Suppl.\  {\bf 82}, 291 (2000)
[hep-ph/9907353]; \\
S.~J.~Brodsky,
eConf {\bf C0309101}, WEPL001 (2003)
[hep-ph/0311355].

\bibitem{Brodsky:2004wx}
S.~J.~Brodsky,
\textit{High energy photon photon collisions at a linear collider},
hep-ph/0404186.


\bibitem{exp}
P.~Achard {\it et al.}  [L3 Collaboration],
Phys.\ Lett.\ B {\bf 536}, 24 (2002)
[hep-ex/0204025];
\\
G.~Abbiendi {\it et al.}  [OPAL Collaboration],
Eur.\ Phys.\ J.\ C {\bf 28}, 45 (2003)
[hep-ex/0209052];
\\
C.~C.~Kuo, S.~C.~Kao, A.~Chen, W.~T.~Chen and S.~Uehara  [Belle
                  Collaboration],
Nucl.\ Phys.\ A {\bf 721}, 817 (2003);
 \\
P.~Achard {\it et al.}  [L3 Collaboration],
Phys.\ Lett.\ B {\bf 571}, 11 (2003)
[hep-ex/0306017];
 \\
T.~Barillari,
Nucl.\ Phys.\ Proc.\ Suppl.\  {\bf 126}, 301 (2004)
[hep-ex/0307066];
 \\
D.~Urner,
AIP Conf.\ Proc.\  {\bf 698}, 566 (2004)
[hep-ex/0309045].


\bibitem{theor}
M.~Diehl, P.~Kroll and C.~Vogt,
Eur.\ Phys.\ J.\ C {\bf 26}, 567 (2003)
[hep-ph/0206288];
\\
P.~Kroll,
Fizika B {\bf 13}, 153 (2004)
[hep-ph/0310327].


\bibitem{Berger:2002vc}
C.~F.~Berger and W.~Schweiger,
Eur.\ Phys.\ J.\ C {\bf 28}, 249 (2003)
[hep-ph/0212066].


\bibitem{Karliner:2001ut}
M.~Karliner,
in
{\it Proc. of the $e^+ e^-$ Physics at Intermediate
Energies Conference }
ed. Diego Bettoni,
eConf {\bf C010430}, W10 (2001)
[hep-ph/0108106].

\bibitem{Farrar:gv}
G.~R.~Farrar, E.~Maina and F.~Neri,
Nucl.\ Phys.\ B {\bf 259}, 702 (1985)
[Erratum-ibid.\ B {\bf 263}, 746 (1986)].


\bibitem{Anselmino:1987gu}
M.~Anselmino, F.~Caruso, P.~Kroll and W.~Schweiger,
Int.\ J.\ Mod.\ Phys.\ A {\bf 4}, 5213 (1989).


\bibitem{Farrar:1988vz}
G.~R.~Farrar, H.~Zhang, A.~A.~Ogloblin and I.~R.~Zhitnitsky,
Nucl.\ Phys.\ B {\bf 311}, 585 (1989).

\bibitem{Karliner:2002nk}
M.~Karliner and S.~Nussinov,
Phys.\ Lett.\ B {\bf 538}, 321 (2002)
[hep-ph/0202234].

\bibitem{Lepage:1980fj}
G.~P.~Lepage and S.~J.~Brodsky,
Phys.\ Rev.\ D {\bf 22}, 2157 (1980);
\\
S.~J.~Brodsky and G.~P.~Lepage,
Adv.\ Ser.\ Direct.\ High Energy Phys.\  {\bf 5}, 93 (1989).

\bibitem{Efremov:1979qk}
A.~V.~Efremov and A.~V.~Radyushkin,
Phys.\ Lett.\ B {\bf 94}, 245 (1980).

\bibitem{Anselmino:1987vk}
M.~Anselmino, P.~Kroll and B.~Pire,
Z.\ Phys.\ C {\bf 36}, 89 (1987).


\bibitem{Jakob:1993th}
R.~Jakob, P.~Kroll, M.~Sch\"urmann and W.~Schweiger,
Z.\ Phys.\ A {\bf 347}, 109 (1993)
[hep-ph/9310227].


\bibitem{prediqu1}
P.~Kroll, M.~Sch\"urmann and W.~Schweiger,
Int.\ J.\ Mod.\ Phys.\ A {\bf 6}, 4107 (1991);
 \\
P.~Kroll, T.~Pilsner, M.~Sch\"urmann and W.~Schweiger,
Phys.\ Lett.\ B {\bf 316}, 546 (1993)
[hep-ph/9305251].

\bibitem{prediqu2}
C.~F.~Berger, B.~Lechner and W.~Schweiger,
Fizika B {\bf 8}, 371 (1999)
[hep-ph/9901338];
 \\
C.~F.~Berger, B.~J\"ager and W.~Schweiger,
Nucl.\ Phys.\ A {\bf 689}, 390 (2001)
[hep-ph/0009295];
 \\
B.~J\"ager and W.~Schweiger,
Nucl.\ Phys.\ A {\bf 711}, 203 (2002)
[hep-ph/0206189];
\\
C.~F.~Berger and W.~Schweiger,
Fizika B {\bf 13}, 74 (2004)
[hep-ph/0309095].


\bibitem{Berger:1999gx}
C.~F.~Berger and W.~Schweiger,
Phys.\ Rev.\ D {\bf 61}, 114026 (2000)
[arXiv:hep-ph/9910509].

\bibitem{Argus}
H.~Albrecht {\it et al.}  [ARGUS collaboration], Z.\ Phys.\ C {\bf
42}, 543 (1989).


\bibitem{Anselmino:vs}
M.~Anselmino, F.~Murgia and F.~Caruso,
Phys.\ Rev.\ D {\bf 42}, 3218 (1990).


\bibitem{Rarita:mf}
W.~Rarita and J.~S.~Schwinger,
Phys.\ Rev.\  {\bf 60}, 61 (1941).


\bibitem{Chernyak:1989}
V.~L.~Chernyak, A.~A.~Ogloblin and I.~R.~Zhitnitsky,
Z.\ Phys.\ C {\bf 42}, 569 (1989)
[Yad.\ Fiz.\  {\bf 48}, 1410 (1988\ SJNCA,48,896-904.1988)].



\bibitem{Bourrely:mr}
C.~Bourrely, J.~Soffer and E.~Leader,
Phys.\ Rept.\  {\bf 59}, 95 (1980);
 \\
E.~Leader,
Cambridge Monogr.\ Part.\ Phys.\ Nucl.\
Phys.\ Cosmol.\  {\bf 15}, 1 (2001).

\bibitem{Echenard}
B.~Echenard, talk at \textit{Photon 2003}, unpublished,
and private
communication.




\bibitem{math}
S.~Wolfram, \textit{The Mathematica Book}, 5th Edition,
Wolfram Media, 2003.

\bibitem{Mertig:1990an}
R.~Mertig, M.~B\"ohm and A.~Denner,
Comput.\ Phys.\ Commun.\  {\bf 64}, 345 (1991).

\end{thebibliography}
\end{document}